\documentclass[aps,prb,preprint,superscriptaddress,showpacs,preprintnumbers,amsmath,amssymb]{revtex4}

\usepackage{graphicx}
\usepackage{longtable}
\usepackage{bm}
\usepackage{dcolumn}

\begin{document}
\title{Crystal structures and magnetic order of
La$_{0.5+\delta}A_{0.5-\delta}$Mn$_{0.5+\epsilon}$Ru$_{0.5-\epsilon}$O$_{3}$
($A=$ Ca, Sr, Ba): Possible orbital glass ferromagnetic state}

\author{E. Granado}
\email{egranado@ifi.unicamp.br}
\affiliation{Instituto de F\'{i}sica ``Gleb Wataghin'', UNICAMP,
Caixa Postal 6165, Campinas, SP, CEP 13083-970, Brazil}
\affiliation{Laborat\'{o}rio Nacional de
Luz S\'{i}ncrotron, Caixa Postal 6192, Campinas, SP, CEP 13084-971, Brazil}

\author{Q. Huang}
\affiliation{NIST Center for Neutron Research, National Institute of Standards and
Technology, Gaithersburg, Maryland 20899}

\author{J. W. Lynn}
\affiliation{NIST Center for Neutron Research, National Institute of Standards and
Technology, Gaithersburg, Maryland 20899}
\affiliation{Center for Superconductivity Research, University of Maryland, College
Park, Maryland 20742}

\author{J. Gopalakrishnan}
\affiliation{Center for Superconductivity Research, University of Maryland, College
Park, Maryland 20742}
\affiliation{Solid State and Structural Chemistry Unit, Indian Institute of Science,
Bangalore 560012, India}

\author{K. Ramesha}
\affiliation{Solid State and Structural Chemistry Unit, Indian Institute of Science,
Bangalore 560012, India}

\begin{abstract}

The crystallographic and magnetic properties of
La$_{0.5+\delta}A_{0.5-\delta}$Mn$_{0.5+\epsilon}$Ru$_{0.5-\epsilon}$O$_{3}$
($A=$ Ca, Sr, Ba) were investigated by means of neutron
powder diffraction. All studied samples show the orthorhombic
perovskite crystal structure, space group P$nma$, with regular (Mn,Ru)O${_6}$ octahedra and no chemical ordering of the
Mn$^{3+}$ and Ru$^{4+}$ ions. Ferromagnetic spin structures were observed below $T_{C}\sim 200-250$ K, with an
average ordered moment of $\sim 1.8-2.0 \mu_{B}$/(Mn,Ru). The observation of long-range ferromagnetism
and the absence of orbital ordering are rationalized in terms a strong Mn-Ru hybridization,
which may freeze the orbital degree of freedom and broaden the $e_{g}$ valence
band, leading to an orbital-glass state with carrier-mediated ferromagnetism.

\end{abstract}


\maketitle

\section{\bf INTRODUCTION}

The large variety of interesting physical phenomena observed in manganese perovskites has
motivated intensive investigations on the phase diagrams of these compounds as functions 
of chemical doping, temperature, pressure, magnetic field, and lattice mismatch, among other variables.
Particularly, a number of recent works has dealt with dilution of transition metal impurities
at the Mn crystallographic site.\cite{Vanitha,Maignan,Martin} The main focus has been on the influence
of such impurities on the relative strength of the ferromagnetic (FM) double exchange interactions against
the competing tendency for charge and orbital ordering of the Mn$^{3+}:e_{g}$ electrons. Less work
has been performed in compounds with large levels ($> 20$ \%) of transition-metal substitution. For 50 \% substitution,
such mixed systems may be classified as double perovskites, with possible transition-metal chemical ordering.
\cite{Anderson}

The particular case of Ru-substitution in manganites appears to yield interesting effects. It is believed that
such ions may be present
in either low-spin Ru$^{4}$ ($4d:t_{2g}^{4}$) or Ru$^{5+}$ ($4d:t_{2g}^{3}$) electronic configurations,
and are substantially more covalent than the $3d$ ions. Previous works indicate that light Ru-substitution
strongly favors the ferromagnetic metallic (FMM) states.\cite{Vanitha,Maignan,Martin,Manoharan}
This is opposite to the general trend observed
for other transition-metal substitutions in manganites, where the $B$-site disorder tends to disrupt the conduction
and exchange paths, suppressing the FMM state.\cite{Vanitha} Despite these interesting results,
relatively few details are presently known about the magnetism of heavily Ru-substituted manganites,
as well as on its possible coupling with the lattice degrees of freedom.

In this work, the crystal and magnetic structures of
La$_{0.5+\delta}A_{0.5-\delta}$Mn$_{0.5+\epsilon}$Ru$_{0.5-\epsilon}$O$_{3}$
($A=$ Ca, Sr, Ba) are investigated by means of high-resolution neutron
powder diffraction (NPD). All samples crystallize in an
orthorhombic perovskite structure, with no chemical ordering of Mn and Ru ions. The (Mn,Ru)O$_{6}$
octahedra are regular, and the Mn-O-Mn bonding angles increase substantially from $A$=Ca to Ba.
Long-range FM structures of (Mn,Ru) spins were observed below $T_{C}\sim 200-250$ K,
with no evidence of coexisting antiferromagnetic order parameters.
The average ordered moment per transition metal ion at 10 K is $\sim 1.8-2.0 \mu_{B}$/(Mn,Ru),
substantially smaller than the $\sim 3 \mu_{B}$/(Mn,Ru) expected for a ferromagnetic ordering of high-spin Mn$^{3+}$ and
low-spin Ru$^{4+}$ moments in an atomistic picture. No lattice anomalies were observed in the studied temperature
interval. These results are interpreted in terms of a strong Mn-Ru hybridization,
which may freeze the orientation of the Mn$^{3+}$ $e_{g}$ orbitals as well as broaden the $e_{g}$ valence
band, possibly leading to carrier-mediated ferromagnetism.

\section{\bf EXPERIMENTAL DETAILS}

Polycrystalline samples with nominal composition La$_{0.5}$Ca$_{0.5}$Mn$_{0.5}$Ru$_{0.5}$O$_{3}$ (LCMR),
La$_{0.5}$Sr$_{0.5}$Mn$_{0.5}$Ru$_{0.5}$O$_{3}$ (LSMR)
and La$_{0.5}$Ba$_{0.5}$Mn$_{0.5}$Ru$_{0.5}$O$_{3}$ (LBMR)
were grown by conventional solid state reaction, as described in ref. [\cite{Ramesha}]. All measurements reported
in this work were taken at the NIST Center for Neutron Research.
The high-resolution NPD measurements were performed on the BT-1 powder diffractometer,
using monochromatic beams with $\lambda =1.5402(1)$ \AA\
and $2.0783(1)$ \AA\, produced by Cu$(311)$ and Ge$(311)$ monochromators, respectively.
Collimations were $15^{\prime }$, $20^{\prime }$, and $7^{\prime }$ arc before and after the
monochromator, and before detectors, respectively. The intensities were measured in steps of 0.05$^{\circ }$
in 2$\theta $ range 3-168$^{\circ }$. The samples were placed into cylindrical vanadium cans. Crystal
and magnetic structure refinements were carried out using the program GSAS.\cite{Larson} The nuclear scattering
amplitudes are $0.827$, $0.490$, $0.702$, $0.525$, $-0.373$, $0.721$, and $0.581$ ($\times 10^{-14}$ m) for La, Ca, Sr, Ba, Mn, Ru, and O,
respectively.\cite{Larson} For LCMR, an impurity phase of LaCa$_{2}$RuO$_{6} $ ($6.5$ \% weight fraction)
with monoclinic crystal structure (P$2_{1}/n$ symmetry)\cite{Battle} was detected, while for LBMR impurity phases
of Ba$_{3}$RuMn$_{2}$O$_{9}$\cite{Frenzen} and BaRuO$_{3}$\cite{Hong} were identified with weight fractions of
5.8 \% and 1.5 \%, respectively. Such impurity phases were included in the refinements for the respective sample.
For LSMR, some Bragg peaks remained unindexed, being ascribed to unidentified impurity phases. The 
intensity of the strongest of such peaks is 5 \% of the strongest peak of the main phase. The relatively poor
fitting thus obtained for LSMR ($\chi^{2}=3.1$) prevented a reliable refinement of some structural parameters
of the main phase for this particular sample. 
 
High-intensity NPD
measurements were performed for LCMR at the BT-7 triple axis spectrometer operated on two-axis mode, using a monochromatic
beam with $\lambda = 2.465(1)$ \AA\ produced by a pyrolytic graphite monochromator. Relaxed collimations
were employed in order to optimize intensity, leading to an instrumental resolution in $2\theta$ of
$\sim 1.0-1.3 ^{\circ}$ full width at half maximum at the angular region of interest to this work. The sample
was placed into an Al cylindrical can to minimize incoherent scattering by the sample holder.
Both high-resolution and high-intensity data were collected at various temperatures in the range 10-300 K, employing
closed-cycle He cryostats. 

\section{\bf RESULTS AND ANALYSIS}

\begin{figure}
\includegraphics[width=1.0 \textwidth]{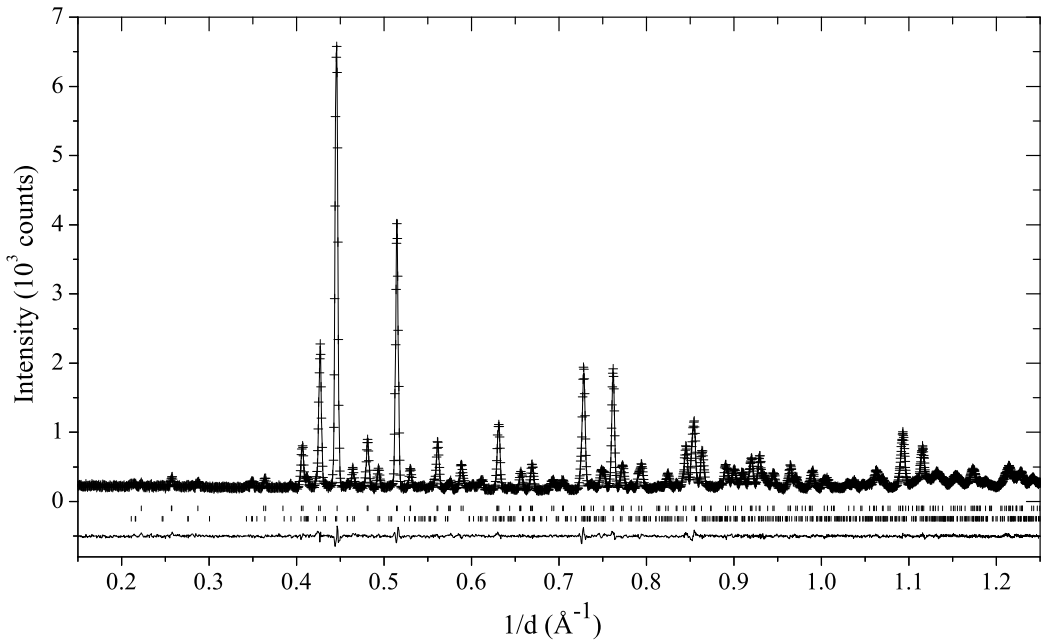}
\caption{\label{ProfileLaCa} Observed (cross symbols) and calculated (solid line) high-resolution neutron diffraction
intensities of La$_{0.5+\delta}$Ca$_{0.5-\delta}$Mn$_{0.5+\epsilon}$Ru$_{0.5-\epsilon}$O$_{3}$ at 300 K, taken
with $\lambda =1.5402(1)$ \AA.
The difference profile is also given. Short vertical lines correspond to Bragg peak positions for the main phase
(upper) and for the LaCa$_{2}$RuO$_{6}$ impurity phase (lower).}
\end{figure}

Figure \ref{ProfileLaCa} shows the observed NPD intensities of LCMR at room-$T$
(symbols). The crystal structure of this compound was refined using an orthorhombic perovskite
model (P$nma$ symmetry), with Mn and Ru ions at the same crystallographic site, as well as
La and Ca ions. The occupancies $\nu$ were refined under the constraints $\nu$(Ca)$+\nu$(Ba)$=1$ and
$\nu$(Mn)$+\nu$(Ru)$=1$. Preliminary refinements showed the oxygen occupancy close to the stoichiometric
value, $\nu$(O)$=3.00(2)$. This parameter was then fixed at 3 in subsequent refinements in order to improve
the stability of the fitting procedure. The calculated NPD intensities using the above model are given in
Fig. \ref{ProfileLaCa} (solid line), yielding good fits to the observed intensities. The crystal structures of LSMR
and LBMR were also refined using the same ionic-disordered model with P$nma$ symmetry
used for LCMR. The refined structural parameters of LCMR and LBMR at 300 K are given in Table \ref{Table1}, together
with selected bond distances and bonding angles. The average $B$-O bond distance  ($B$=Mn,Ru) assumes
nearly the same value, $\sim 1.99$\ \AA, for both LCMR and LBMR. Also,
the $B$O$_{6}$ octahedra are nearly regular, since the $B$-O bond distances show only
a small distribution in both samples, within $\sim 0.02$\ \AA. The unit cell volume shows a significant expansion
for LBMR with respect to LCMR. This is accomplished by a rotation of the $B$O$_{6}$ octahedra in order to
increase the $B-$O$-B$ bonding angle, thus expanding the average Ba$-$O bond distances with respect to Ca$-$O. We note
that the $B-$O1$-B$ angle increases from $155.32(12)^{\circ}$ for LCMR to $163.2(3)^{\circ}$ for LBMR, while
the $B-$O2$-B$ angle experiences a more significant change, from $155.51(7)^{\circ}$ for LCMR to $175.5(4)^{\circ}$
for LBMR. 

\begin{figure}
\includegraphics[width=0.5 \textwidth]{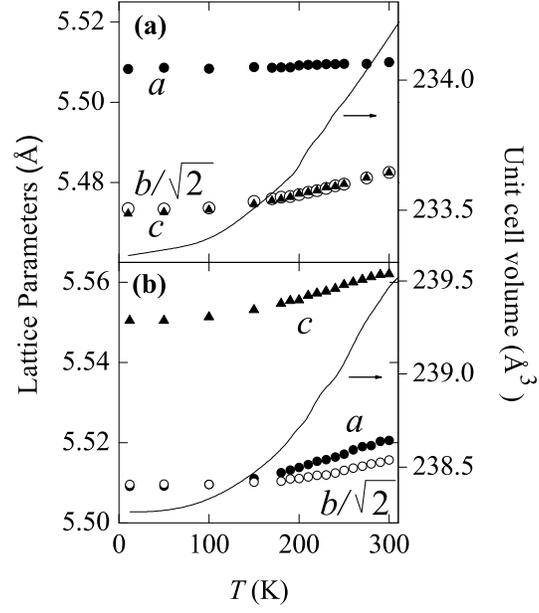}
\caption{\label{lattpar} Temperature dependence of the lattice parameters {\it a} (filled
circles), $b/\sqrt{2}$ (open circles), and {\it c} (filled tringles), and unit cell
volume (solid line) of
(a) La$_{0.5+\delta}$Ca$_{0.5-\delta}$Mn$_{0.5+\epsilon}$Ru$_{0.5-\epsilon}$O$_{3}$, and
(b) La$_{0.5+\delta}$Sr$_{0.5-\delta}$Mn$_{0.5+\epsilon}$Ru$_{0.5-\epsilon}$O$_{3}$.}
\end{figure}

\begin{figure}
\includegraphics[width=0.5 \textwidth]{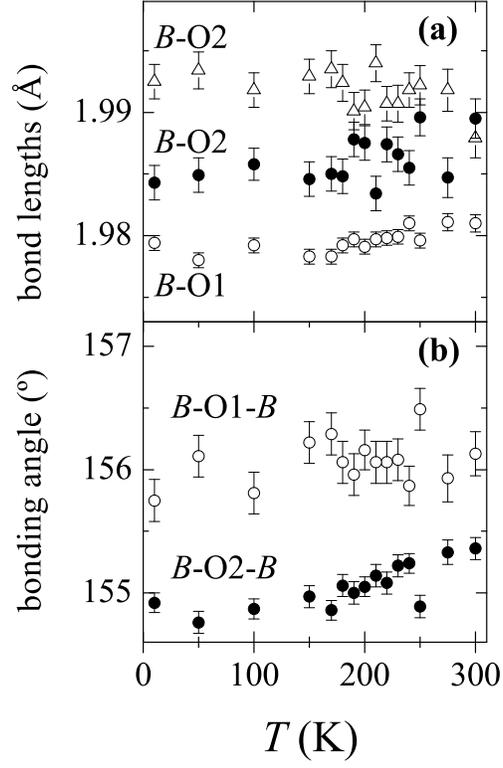}
\caption{\label{bonds} Temperature-dependence of (a) $B$-O bond lengths ($B=$ Mn, Ru)
and (b) $B-$O$-B$ bonding angles for
La$_{0.5+\delta}$Ca$_{0.5-\delta}$Mn$_{0.5+\epsilon}$Ru$_{0.5-\epsilon}$O$_{3}$.}
\end{figure}

Figures \ref{lattpar}(a) and \ref{lattpar}(b) show the temperature-dependence of the lattice parameters $a$, $b$, and
$c$ and unit-cell volume
for LCMR and LSMR, respectively. The lattice parameters show a conventional contraction with decreasing $T$, with
no detectable anomaly in the studied temperature interval. The temperature-dependencies of $B-$O bond distances and
$B-$O$-B$ bonding angles for LCMR are given in Fig. \ref{bonds}(a) and \ref{bonds}(b), respectively. The
$B-$O bond lenghts remains nearly constant over the studied $T$-interval, within $\sim 0.001$ \AA. 

\begin{figure}
\includegraphics[width=0.5 \textwidth]{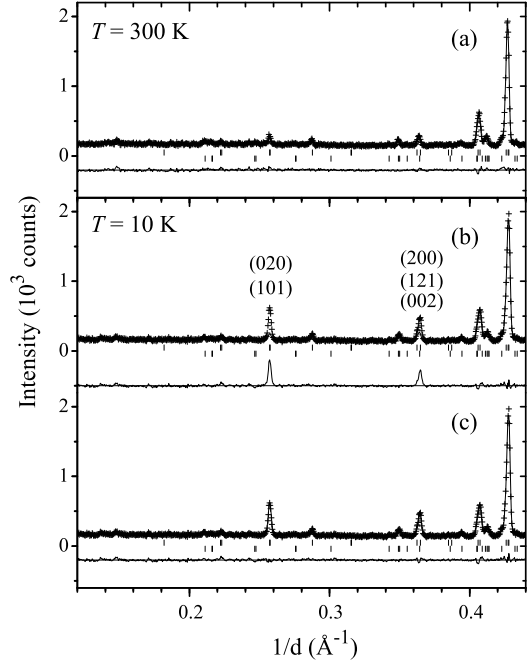}
\caption{\label{lowangle} Low angle portion of the observed high-resolution neutron diffraction intensities of
La$_{0.5+\delta}$Ca$_{0.5-\delta}$Mn$_{0.5+\epsilon}$Ru$_{0.5-\epsilon}$O$_{3}$ (cross symbols) at (a) 300 K
and (b,c) 10 K, taken with
the Ge(311) monochromator. Calculated intensities (solid lines) are also given, for a nuclear-only model (a,b), and 
for a nuclear+ferromagnetic model for the Mn and Ru spins (c). The difference profiles are also shown.
Short vertical lines correspond to Bragg peak positions for the main phase
(upper) and for the LaCa$_{2}$RuO$_{6}$ impurity phase (lower).}
\end{figure}

Figures \ref{lowangle}(a) and \ref{lowangle}(b) show a low-angle portion of the high-resolution
neutron powder profile of LCMR, taken with Ge(311) monochromator at $T$ =
300 K and $T$ = 10 K, respectively, and the calculation according to the
lattice model described above. Clearly, additional contributions to the
intensities of the (101)/(020) and (200)/(121)/(002) Bragg peaks are observed at
$T=10$ K (see difference curve in Fig. \ref{lowangle}(b)), arising from the ordering of
the Mn(Ru) spins. The widths of the FM Bragg peaks are found to be instrumental-only,
thus FM domains are larger than $\sim 500$ \AA\ for all studied samples. Figure \ref{lowangle}(c)
shows a comparison between the observed
profile at $T$ = 10 K and a model including a ferromagnetic ordering for the Mn and Ru spins, yielding good
agreement with each other.  

\begin{figure}
\includegraphics[width=0.5 \textwidth]{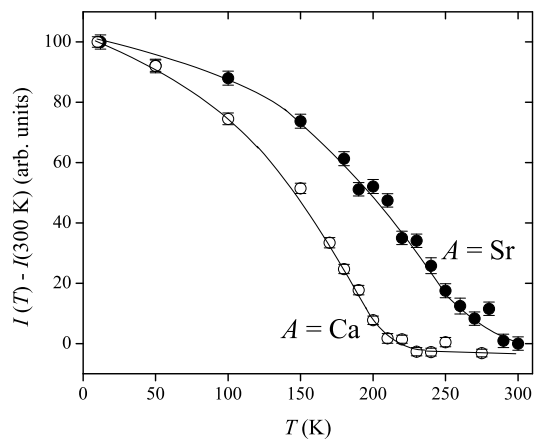}
\caption{\label{magintensity} Temperature dependence of the summed integrated intensity of the
(101), (020), (200), (002), and (121) Bragg reflections which have significant contributions from a
ferromagnetic spin arrangement of Mn and Ru spins. Solid curves are guides to the eye.}
\end{figure}

Figure \ref{magintensity} shows the temperature-dependence of the summed integrated
intensity of the (101), (020), (200), (002), and (121) NPD Bragg reflections with significant FM contributions
for LCMR and LSMR.
Ferromagnetic ordering is observed below $T_{C} \sim 200$ K for LCMR and $T_{C} \sim 250$ K for LSMR,
consistent with $dc$-magnetization measurements previously performed in these samples.\cite{Ramesha}
Table \ref{Table2} summarizes the magnetic properties of the studied samples, obtained from our refinements
using high-resolution NPD data except where noted otherwise.

\begin{figure}
\includegraphics[width=0.5 \textwidth]{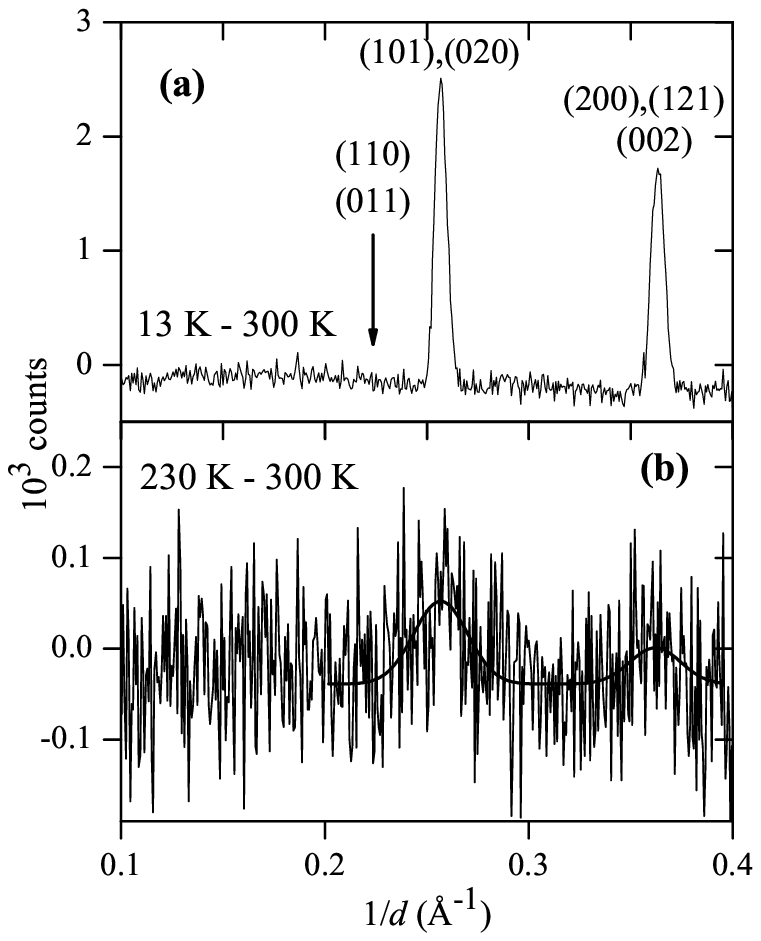}
\caption{\label{bt7} High-intensity difference neutron profiles for 
La$_{0.5+\delta}$Ca$_{0.5-\delta}$Mn$_{0.5+\epsilon}$Ru$_{0.5-\epsilon}$O$_{3}$ : (a) $I$(13 K$)-I$(300 K),
and (b) $I$(230 K$)-I$(300 K), highlighting the magnetic scattering well below and slightly above $T_{C}$,
respectively. The smooth solid line in (b) is a two-Gaussian fit to the experimental data.}
\end{figure}

In order to search for weak magnetic signals not observable in high-resolution measurements,
high-intensity energy-integrated NPD measurements were performed for LCMR. Figure \ref{bt7}(a)
shows a portion of the neutron scattering intensities at 10 K ($<<T_{C}$) subtracted by the intensities at 300 K
($>>T_{C}$). Besides the magnetic Bragg peaks already seen in the high-resolution measurements (see 
Figs. \ref{lowangle}(b) and \ref{lowangle}(c)), associated with a FM arrangement of (Mn,Ru) spins, no evidence for
additional magnetic sublattices and/or competing magnetic structures or correlations were observed at low temperatures.
Particularly, a chemical ordering of Mn and Ru ions in a double perovskite structure with either ferro or ferrimagnetic
spin arrangement at low temperatures would lead to magnetic contributions to the (110) and (011) Bragg peaks (arrow
in Fig. \ref{bt7}(a)). The absence of such a contribution in Fig. \ref{bt7}(a) demonstrates the absence of a significant
volumetric fraction of the sample showing an ordered double perovskite crystal structure and contributing to the
magnetism down to 10 K. In addition, no evidence of magnetic ordering in the impurity phases of this sample was
observed. Figure \ref{bt7}(b)
shows a similar measurement at 230 K, therefore slightly above $T_{C} \sim 200$ K, again subtracted by the scattering
at 300 K. Significant short-range ferromagnetic correlations are observed at this temperature. From the widths of
the broad structures centered at the positions where the FM peaks develop at low temperatures (solid line in
Fig. \ref{bt7}(b)), a magnetic correlation length of 35(10) \AA\ is found for LCMR at 230 K.

\section{\bf DISCUSSION}

Our results indicate that a FM state with a relatively low spontaneous moment ($\sim 1.8-2.0 \mu_{B}/$(Mn,Ru))
is realized 
for La$_{0.5+\delta}A_{0.5-\delta}$Mn$_{0.5+\epsilon}$Ru$_{0.5-\epsilon}$O$_{3}$ ($A=$ Ca, Sr, Ba). No other 
coexisting magnetic sublattices were detected. In addition, this system does not display lattice anomalies
that might be associated with orbital ordering, or volumetric changes at $T_{C}$ due to an increase of the
kinetic energy of $d$-electrons in the FM phase, as observed in FM manganites without Ru-substitution. We note
that the system studied here is not metallic and does not show an appreciable change in conductivity at the onset
of the FM phase.\cite{Ramesha}

As a first step to rationalize the above observations, a discussion of the oxidation states of the Mn and Ru ions
is worthwhile. Two possibilities arise from the charge neutrality condition for the formula unit: (i) Mn$^{2+}-$Ru$^{5+}$,
and (ii) Mn$^{3+}-$Ru$^{4+}$. Based on our
results, we argue against (i). In fact, the large differences between Mn$^{2+}$ and Ru$^{5+}$ in either
electrostatic charge ($3e$) and ionic radius (0.265 \AA)\cite{Shannon} would most likely lead to a chemically-ordered double
perovskite structure.\cite{Anderson} Given the large differences between the nuclear neutron scattering factors of
Mn and Ru (with opposite signs, see Section II above), a significant volume fraction of such an ordered structure would
be readly identified in our NPD profiles, which was not the case (see Fig. \ref{ProfileLaCa}).
Even the presence of a significant volumetric
fraction of short-range chemically ordered domains\cite{Ramesha} is not consistent with our measurements. In fact, the
magnetic state at low-temperatures associated with such hypothetical clusters (either ferro- or ferrimagnetic) would
lead to additional structures in our low-temperature NPD profiles, centered at positions such as (110)/(011) in
Fig. \ref{bt7}. As a final argument against a significant presence of Mn$^{2+}$/Ru$^{5+}$ pairs, the expected average
(Mn,Ru)-O distance for this configuration is 2.10 \AA,\cite{Shannon} much larger than
the observed values, $\sim 1.99$ \AA\ (see Table \ref{Table1}). We note that the average (Mn,Ru)-O
distance expected for Mn$^{3+}$/Ru$^{4+}$ is 2.03 \AA,\cite{Shannon} fairly close to the experimental values. The remaining
discrepancy may be partly ascribed to a possible presence of a small fraction of Mn$^{4+}$ or Ru$^{5+}$ ions caused by 
off-stoichiometry effects ($\delta, \epsilon \neq 0$). We should mention that superlattice Bragg reflections indicative of
a chemically-ordered double perovskite structure have been observed by electron diffraction for these same samples
studied here.\cite{Ramesha} Also, the crystal structure of La$_{0.5+x}$Sr$_{0.5-x}$Mn$_{0.5}$Ru$_{0.5}$O$_{3}$ was found
to show a transition from random to ordered distribution of Mn and Ru ions at $x=0.0$.\cite{Fang}
Nonetheless, our NPD results show that a chemically-ordered state is not representative of the bulk of the LCMR, LSMR,
and LBMR samples studied in this work. A chemically-disordered perovskite phase for LSMR was also reported by
Horikubi {\it et al.}\cite{Horikubi}

In view of the above, the
La$_{0.5+\delta}A_{0.5-\delta}$Mn$_{0.5+\epsilon}$Ru$_{0.5-\epsilon}$O$_{3}$ ($A=$ Ca, Sr, Ba) compounds may be described
as a random distribution of Mn$^{3+}$ and Ru$^{4+}$ ions in the $B$-site of a perovskite structure. It is remarkable
that a long-range FM state can survive to such a degree of random Ru substitution at the Mn site. 
We base our discussion upon a scenario proposed by Martin {\it et al.}\cite{Martin}
to explain the rapid quenching of the orbital-ordered state and the development of a FMM phase in Sm$_{0.5}$Ca$_{0.5}$MnO$_{3}$
with relatively small levels of Ru-substitution. According to this, the spatially-extended character
of the Ru $4d$ orbitals would lead the $4d: e_{g}$ states to participate to the band formation and contribute to make it
broader, producing a FM coupling mechanism between mixed-valent Mn and Ru$^{4+}$ ions based upon an enhancement of
the double exchange interaction. We note that a broadening of the $e_{g}$ band by Mn/Ru hybridization would lead
to a FM coupling between Mn and Ru only if the $e_{g}$ band is partially filled, i.e., the $e_{g}$ density of
states must be finite at the Fermi level. In the present case, our NPD measurements suggest that the Mn oxidation state
is close to Mn$^{3+}$ (see discussion above), with one $e_{g}$ electron per Mn ion. In this case, the likelyhood of
$e_{g}$ charge carriers increases with the ratio of the
$e_{g}$ bandwidth by the energy splitting. The presence of Ru ions is believed to increase the $e_{g}$ bandwidth by
Mn/Ru hybridization, and may decrease the Jahn-Teller splitting, since the local distortion of the Mn$^{3+}$O$_{6}$
octahedra might be smoothed out by the chemical disorder. It is therefore not unreasonable to assume 
the presence of a significant $e_{g}$ density of states at the Fermi level in
La$_{0.5+\delta}A_{0.5-\delta}$Mn$_{0.5+\epsilon}$Ru$_{0.5-\epsilon}$O$_{3}$ ($A=$ Ca, Sr, Ba). If this assumption
is correct, the above carrier-mediated mechanism for the FM coupling may be valid. We should mention that 
transport measurements do not show metallic behavior below $T_{C}$, although resistivity was found to be significantly
low with a thermal behavior not characteristic of conventional semiconductors.\cite{Ramesha}
In fact, a presumably small carrier density combined with the strong chemical disorder of
this system would make a metallic state highly unlikely even in the FM phase, due to Anderson localization.
More direct investigations of the electronic structure of this system are required in order to confirm or dismiss
the validity of the above scenario. Notice that this is consistent with a previous report evidencing the presence of double
exchange interactions in La$_{0.7}$Pb$_{0.3}$Mn$_{1-x}$Ru$_{x}$O$_{3}$ up to large levels of Ru substitution ($x \sim 0.4$).
\cite{Manoharan} We should mention that a FM Mn$^{3+}-$Ru$^{4+}$ superexchange interaction\cite{MartinEuro}
is possibly another important ingredient that may lead to the robust ferromagnetism found in
La$_{0.5+\delta}A_{0.5-\delta}$Mn$_{0.5+\epsilon}$Ru$_{0.5-\epsilon}$O$_{3}$ ($A=$ Ca, Sr, Ba). On the other hand,
the same values of $T_{C}$ found for $A=$Ca and $A=$Ba (see Table \ref{Table2}) despite the large differences
in the (Mn,Ru)-O-(Mn,Ru) superexchange angles (see Table \ref{Table1}) are intriguing, and suggest that superexchange
alone is not sufficient to understand the magnetic properties of this system.

The ordered moments found at low temperatures ($\sim 1.8-2.0 \mu_{B}$/(Mn,Ru), see Table \ref{Table2}) are significantly
lower than the $3 \mu_{B}$/(Mn,Ru) expected for a FM arrangement of high-spin Mn$^{3+}$ ($4 \mu_{B}$) and low-spin Ru$^{4+}$
($2 \mu_{B}$) moments in a spin-only atomistic picture. We first note that the Ru $4d$ states are highly hybridized
with O $2p$, which may reduce the Ru moments. We mention that SrRuO$_{3}$ and Sr$_{4}$Ru$_{3}$O$_{10}$ are metallic
ferromagnets with saturated magnetic moments well below $2 \mu_{B}$/Ru$^{4+}$.\cite{Randall,Crawford}
Secondly, the random distribution of Mn and Ru ions in the lattice might lead to some sites with frustrated
exchange interactions in the FM lattice, leading to local spin reversal and/or loss of spin collinearity.
A combination of the two effects above may lead to the reduced FM moments observed by NPD. 

In transition-metal compounds with a significant density of degenerate
$t_{2g}$ and $e_{g}$ orbitals, strong tendencies for orbital-ordering (OO) have been identified.\cite{Tokura} The spin
states associated with OO are generally antiferromagnetic, therefore OO
and ferromagnetism (satisfying double-exchange interactions) tend to be competing ground states. Such a competition is believed
to be a decisive ingredient leading to the rich phase diagram observed in manganites, in addition to tendencies for
charge ordering for specific concentrations of $e_{g}$ electrons, which also favors OO.
In the present case, the FM coupling is believed to arise from hybridization between Mn$^{3+}$ and Ru$^{4+}$ $e_{g}$
states, leading to a presumably small but significant carrier density in the $e_{g}$ band. One might suspect
that tendencies for OO should overcome the FM coupling
in La$_{0.5+\delta}A_{0.5-\delta}$Mn$_{0.5+\epsilon}$Ru$_{0.5-\epsilon}$O$_{3}$ ($A=$ Ca, Sr, Ba),
since both Mn$^{3+}$ and Ru$^{4+}$ ions show degenerate $d$ orbitals. However, the FM ground state actually observed
in this system, with no cooperative distortions of the (Mn,Ru)O$_{6}$ octahedra (see Table \ref{Table1}) 
and/or lattice anomalies that might be associated with OO (see Figs. \ref{lattpar} and \ref{bonds}),
contradicts such an expectation, indicating that Ru-substitution actually inhibits OO.
In fact, previous studies indicate that even a light Ru substitution quenches orbitally-ordered states in
half-doped manganites.\cite{Martin} The most likely explanation for this tendency stems from the spatially-extended
character of the Ru $4d$ states: the Mn$^{3+}$
$e_{g}$ orbitals would have a tendency to point along the neighboring Ru ions, in order to take full advantage of the
strong Mn-Ru hybridization.\cite{Martin} Notice that each Mn$^{3+}$ ion has two Mn/Ru neighbors along each
binding direction. Thus, each Mn$^{3+}$ $e_{g}$ orbital would point to the direction with most Ru ions. 
This effect may freeze the orbital degree of freedom of Mn$^{3+}$ $e_{g}$ and possibly
Ru$^{4+}$ $t_{2g}$ electrons. In a crystal structure with a random distribution of Ru and Mn ions (see Section III),
long-range OO would be prevented, leading to an orbital glass state.

In conclusion, our NPD results on
La$_{0.5+\delta}A_{0.5-\delta}$Mn$_{0.5+\epsilon}$Ru$_{0.5-\epsilon}$O$_{3}$ ($A=$ Ca, Sr, Ba) show a
perovskite crystal structure with no chemical ordering of Mn$^{3+}$ and Ru$^{4+}$ ions. No evidence of long-range
orbital ordering was observed, possibly due to a freezing of the orbital degree of freedom caused by a significant
Mn-Ru hybridization. A long-range FM spin structure was observed. This result was ascribed to the spatially-extended
character of Ru $4d$ levels,
contributing to broaden the $e_{g}$ band and leading to the presence of a small but finite density of $e_{g}$ states
at the Fermi level. Such states may lead to carrier-mediated FM coupling. In this context, the non-metallic conductivity
observed in  this system\cite{Ramesha} was ascribed to Anderson localization caused by the large Ru/Mn chemical disorder.

\section{Acknowledgements}

Work at the University of Maryland was supported by the NSF-MRSEC, DMR 00-80008.
This work was also supported by Fapesp and CNPq, Brazil, and DST, New Delhi, India.

\begingroup
\squeezetable
\begin{table*}
\caption{\label{Table1}Results of Rietveld-refinements of neutron powder diffraction data collected at 300 K
for La$_{0.5+\delta}A_{0.5-\delta}$Mn$_{0.5+\epsilon}$Ru$_{0.5-\epsilon}$O$_{3}$
($A=$ Ca, Ba). Refinements were carried out in space group P$nma$ (\#62). Errors in parentheses are statistical
only, and represent one standard deviation}
\begin{ruledtabular}
\begin{tabular}{c c c} 
$A$ & Ca & Ba \\
\hline \\
$a$ (\AA) & 5.5099(2) & 5.5975(5)\\
$b$ (\AA) & 7.7553(3) & 7.9163(8)\\
$c$ (\AA) & 5.4819(3) & 5.6197(8)\\
& & \\
La$_{0.5+\delta}A_{0.5-\delta} (x,1/4,z)$ & & \\
$\delta$ & 0.056(12) & 0.04(3) \\
$x$ & 0.0322(2) & -0.0069(14) \\
$z$ & 0.0067(4) & -0.0067(12) \\
$U_{iso}$ (\AA$^{2}$) & 0.0120(4) & 0.0101(9)\\
& & \\
Mn$_{0.5+\epsilon}$Ru$_{0.5-\epsilon} (1/2,0,0)$ & & \\
$\epsilon$ & 0.081(3) & 0.024(3) \\
$U_{iso}$ (\AA$^{2}$) & 0.009(2) & 0.009$^{a}$ \\
& & \\
O1 $(x,1/4,z)$ & & \\
$x$ & 0.4833(3) & 0.502(2) \\
$z$ & -0.0755(4) & 0.0521(9) \\
$U_{11},U{33}$ (\AA$^{2}$) & 0.0183(7) & 0.034(2) \\
$U_{22}$ (\AA$^{2}$) & 0.0068(10) & 0.000(3)\\
& & \\
O2 $(x,y,z)$ & & \\
$x$ & 0.2891(2) &  0.252(3)\\
$y$ & 0.0387(2) & -0.0099(7)\\
$z$ & 0.2872(2) & 0.252(3)\\
$U_{11},U_{33}$ (\AA$^{2}$) & 0.0147(3) & 0.025(2)\\
$U_{22}$ (\AA$^{2}$) & 0.0131(6) & 0.018(3)\\
$U_{13}$ (\AA$^{2}$) & -0.0013(6) & 0.012(2)\\
\hline \\
(Mn,Ru)-O1 (\AA) & 1.9847(5) & 2.0006(7)\\
(Mn,Ru)-O2 (\AA) & 1.9796(11) & 1.9845(3)\\
(Mn,Ru)-O2 (\AA) & 1.9970(12) & 1.9846(4)\\
$<$(Mn,Ru)-O$>$ (\AA) & 1.9871(6) & 1.9899(3)\\
& & \\
(Mn,Ru)-O1-(Mn,Ru) ($^{\circ}$) & 155.32(12) & 163.2(3)\\
(Mn,Ru)-O2-(Mn,Ru) ($^{\circ}$) & 155.51(7) & 175.4(4)\\

\hline \\

$Rp (\%)$ & 3.2 & 5.1\\
$wRp (\%)$ & 3.8 & 6.3\\
$\chi^{2}$ & 1.06 & 1.44\\
\end{tabular}
\end{ruledtabular}
$^{a}$ kept fixed at the same value found for $A=$Ca to avoid instabilities in the fitting procedure. 
\end{table*}
\endgroup

\begingroup
\squeezetable
\begin{table*}
\caption{\label{Table2} Summary of the magnetic properties of 
La$_{0.5+\delta}A_{0.5-\delta}$Mn$_{0.5+\epsilon}$Ru$_{0.5-\epsilon}$O$_{3}$ ($A=$ Ca, Sr, Ba) samples, obtained
from the refinements of neutron powder diffraction data.}
\begin{ruledtabular}
\begin{tabular}{c c c c} 
$A$ & Ca & Sr & Ba \\
\hline \\
Magnetic structure & ferromagnetic & ferromagnetic & ferromagnetic \\
Ordered moments at 10 K & $1.77(2)\mu_{B}$/(Mn,Ru) & $1.83(5) $/(Mn,Ru) & $1.96(4)$/(Mn,Ru) \\
$T_{C}$ & $\sim 200$ K & $\sim 250$ K & $\sim 200$ K$^{a}$ \\
\end{tabular}
\end{ruledtabular}
$^{a}$ taken from ref. \cite{Ramesha}
\end{table*}
\endgroup

\newpage 

\newpage

\begin{references}

\bibitem{Vanitha} P.V. Vanitha, A. Arulraj, A.R. Raju, and C.N.R. Rao, C. R.
Acad. Sci., Ser. IIc: Chim {\bf 2}, 595 (1999).

\bibitem{Maignan} A. Maignan, C. Martin, M. Hervieu, and B. Raveau, Solid
State Commun. {\bf 117}, 377 (2001).

\bibitem{Martin} C. Martin, A. Maignan, M. Hervieu, C. Autret, B. Raveau,
and D.I. Khomskii, Phys. Rev. B {bf 63}, 174402 (2001).

\bibitem{Anderson}  for a review, see M.T. Anderson, K.B. Greenwood, G.A.
Taylor, and K.R. Poeppelmeier, Prog. Solid St. Chem. {\bf 22}, 197 (1993).

\bibitem{Manoharan} S.S. Manoharan, R.K. Sahu, M.L. Rao, D. Elefant, and
C.M. Schneider, Europhys. Lett. {\bf 59}, 451 (2002).

\bibitem{Ramesha}  K. Ramesha, V. Thangadurai, D. Sutar, S.V. Subramanyam,
G.N. Subbanna, and J. Gopalakrishnan, Mater. Research Bull. {\bf 35}, 559
(2000).


\bibitem{Larson}  C. Larson and R.B. Von Dreele, Los Alamos National
Laboratory Report No. LAUR086-748, 1990 (unpublished).

\bibitem{Battle} P.D. Battle, J.B. Goodenough, and R. Price, J. Sol. State
Chem. {\bf 46}, 234 (1983).

\bibitem{Frenzen} S. Frenzen and H. Mueller-Buschbaum, Zeitschrift
fuer Naturforschung {\bf 50}, 585 (1995).

\bibitem{Hong} S.-T. Hong and A.W. Sleight, J. Sol. State Chem. {\bf 128}, 251 (1997).

\bibitem{Shannon} R.D. Shannon, Acta Cryst. {\bf A32}, 751 (1976).

\bibitem{Fang} M. Fang, M. Kato, K. Yoshimura, and K. Kosuge, J. All. Compounds {\bf 317-318}, 136 (2001).

\bibitem{Horikubi} T. Horikubi, T. Mori, H. Nonobe, and N. Kamegashira, J. All. Compounds {\bf 289},
42 (1999).

\bibitem{MartinEuro} C. Martin, A. Maignan, M. Hervieu, B. Raveau, and J. Hejtmanek, Eur. Phys. J B {\bf 16},
469 (2000).

\bibitem{Randall} J.J. Randall and R. Ward, J. Am. Chem. Soc. {\bf 81}, 2629 (1959).

\bibitem{Crawford} M.K. Crawford, R.L. Harlow, W. Marshall, Z. Li, G. Cao, R.L. Lindstrom, Q. Huang, and J.W. Lynn,
Phys. Rev. B {\bf 65}, 214412 (2002).

\bibitem{Tokura} for a review, see Y. Tokura and N. Nagaosa, Science {\bf 288}, 462 (2000).

\end{references}
\end{document}